\documentclass[lettersize,journal]{IEEEtran}
\PassOptionsToPackage{hyphens}{url}

\usepackage{makecell}
\usepackage{longtable}
\usepackage{booktabs}
\usepackage{float}
\usepackage{amsmath,amsfonts}
\usepackage{algorithmic}
\usepackage{algorithm}
\usepackage{array}
\usepackage[caption=false,font=normalsize,labelfont=sf,textfont=sf]{subfig}
\usepackage{textcomp}
\usepackage{stfloats}
\usepackage{url}
\usepackage{verbatim}
\usepackage{graphicx}
\usepackage{cite}
\usepackage{multirow} 
\usepackage[colorlinks,linkcolor=red,anchorcolor=green,citecolor=blue,urlcolor=black]{hyperref}
\usepackage{cleveref}

\usepackage[table]{xcolor}
\crefname{figure}{fig}{figures}
\Crefname{figure}{Fig}{Figures}
\begin{document}

\title{Agentic AI-Enabled Solar-Powered High-Altitude Platforms for Sustainable SAGINs}

\author{Haoxiang Luo, Bang Huang, Mohamed-Slim Alouini,~\IEEEmembership{Fellow,~IEEE}
 
\thanks{H. Luo, B. Huang, and M. S. Alouini are with the Computer, Electrical and Mathematical Science and Engineering (CEMSE) Division in King Abdullah University of Science and Technology (KAUST), Thuwal 6900, Makkah Province, Saudi Arabia. (emails: \{haoxiang.luo, bang.huang, slim.alouini\}@kaust.edu.sa).
}

}



\maketitle

\begin{abstract}

Space-Air-Ground Integrated Networks (SAGINs) can extend connectivity, but their communication, computing, and platform operations create tightly coupled energy demands. Solar-powered High-Altitude Platforms (HAPs) offer a promising middle layer by combining persistent regional coverage, renewable-energy harvesting, and onboard computing. However, realizing this potential requires more than optimizing individual links or processors, as radio transmission, task execution, backhaul use, and battery preservation share a common energy budget. Therefore, we introduce a HAP-native Agentic AI framework. It continuously perceives communication, computing, energy, mobility, and mission states; invokes quantitative tools for prediction and verification; and coordinates executable actions through a closed control loop. Then, a multi-timescale design separates fast radio control from task orchestration and long-term energy planning. Furthermore, a disaster-recovery case study illustrates how the framework responds to backhaul congestion, traffic surges, and declining solar generation, improving energy efficiency, task completion, and latency over other baselines. We finally identify trustworthy control, collaborative multi-HAP orchestration, and digital-twin-assisted lifelong adaptation as key steps toward deployable, sustainable, and resilient SAGIN intelligence.

\end{abstract}


\section{Introduction} \label{sec-I}

\IEEEPARstart{S}{pace}-Air–Ground Integrated Networks (SAGINs) are expected to extend communication and computing services beyond the geographical and economic limits of conventional terrestrial infrastructure \cite{luo2025convergence}. By integrating terrestrial base stations with low-altitude drones, High-Altitude Platforms (HAPs), and satellite constellations, a SAGIN can provide wide-area connectivity, resilient backhaul, remote sensing, distributed inference, and edge computing for disaster response, intelligent transportation, and environmental monitoring \cite{wang2025toward}. 

This expanded capability, however, introduces a fundamental sustainability challenge. A SAGIN consumes energy not only when transmitting information but also when executing computation tasks and maintaining non-terrestrial platforms. Radio-frequency chains, phased-array beamforming, inter-satellite or air-to-space links, task migration, onboard CPUs and AI accelerators, cooling, navigation, and station keeping all compete for limited energy. The problem becomes particularly acute for non-terrestrial nodes because the energy consumed for communication and computing cannot be separated from the energy required to keep the platform operational. And often, due to their limited size, drones are unable to carry large-capacity batteries \cite{huang2022drone}.
A terrestrial network that addresses traffic growth through dense base-station deployment increases grid demand and site construction. Therefore, a sustainable SAGIN should not merely minimize the energy of an individual communication link. It must determine where, when, and by which platform each communication or computing task should be performed, subject to renewable-energy availability and service requirements.

Within this context, solar-powered HAPs occupy a particularly favorable operating point. A HAP typically operates in the stratosphere, between low-altitude aerial networks and satellite systems, and can remain approximately stationary relative to a service region \cite{wang2025toward}. Its large upper surface can accommodate photovoltaic arrays, while onboard batteries store excess daytime energy for night operation. The practical feasibility of this concept has already been demonstrated. For example, the Zephyr stratospheric aircraft\footnote{https://www.aaltohaps.com/zephyr-sets-world-record-for-longest-continuous-flight-flying-67-days-in-stratosphere/} is entirely solar powered and achieved 67 continuous days and nights of stratospheric flight in 2025. Airbus reports\footnote{https://www.airbus.com/en/products-services/defence/uas/zephyr} that a Zephyr equipped with a connectivity payload can provide direct low-latency service over an area of up to approximately 7,500 km$^2$. These figures demonstrate that solar-powered HAPs are no longer only conceptual relay platforms; they are evolving into persistent regional communication infrastructures.


However, in SAGINs, deploying communication and computing functions on solar-powered HAPs still faces four interrelated challenges: \emph{1) Highly dynamic SAGIN network topology} includes satellites, aircraft, and constantly evolving ground traffic hotspots \cite{luo2025convergence}. This requires HAPs to continuously make decisions on link activation and switching; \emph{2) Tight coupling among different performances} includes communication latency, computing efficiency, and energy consumption. This makes local optimization efficient but difficult to achieve global sustainability; \emph{3) High heterogeneity of SAGIN tasks and devices} requires fine coordination of task deployment, model accuracy, and intermediate data routing \cite{luo2025convergence}; \emph{4) Solar energy introduces day-night causality} because propulsion and communication share the harvested-energy budget \cite{javed2023interdisciplinary}. Although their joint optimization improves energy utilization, maximizing instantaneous throughput may still interrupt services after sunset. Therefore, sustainable operation requires long-term service orchestration.

Traditional solutions are insufficient for this environment. Mathematical optimization can provide high-quality solutions when channel, task, mobility, and energy models are well-known. However, repeatedly reformulating and solving large mixed-integer problems is difficult amid rapid topology changes \cite{yao2025multi}. Deep Reinforcement Learning (DRL) can learn resource-allocation policies but is generally trained for a predefined state space, action space, and reward function. It can behave unpredictably when mission objectives, devices, or network layers change. A standalone Large Language Model (LLM) can interpret high-level objectives and explain possible actions \cite{luo2026trustworthy}, but without persistent perception, quantitative tools, execution interfaces, memory, and feedback, it remains a recommendation engine rather than a network controller.

 \begin{figure*}[!t]
\centering
 \includegraphics[width=6 in]{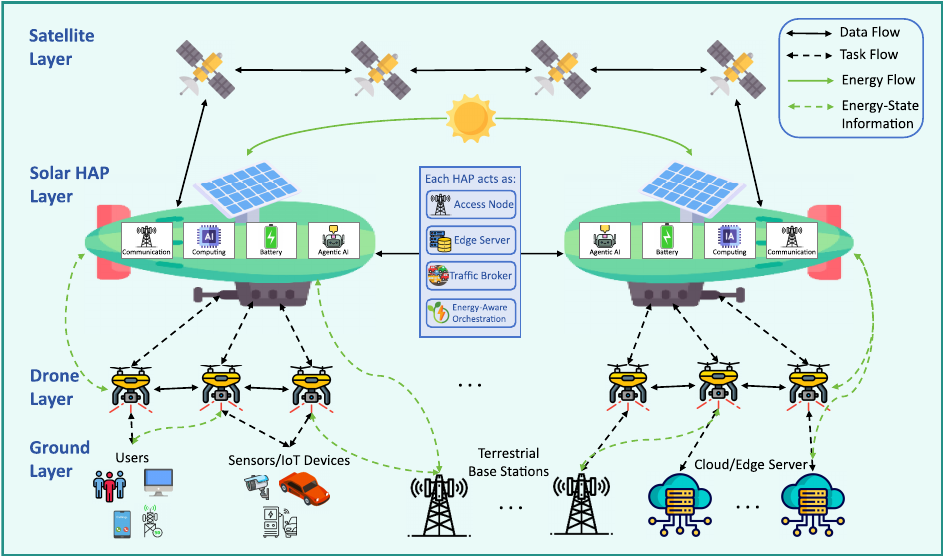}
   \caption{Solar-Powered HAP-Centric SAGIN Architecture. The upper layer contains satellites; the middle layer contains two solar HAPs with communication, computing, battery, and Agentic AI payloads; the lower air layer contains drones; and the ground layer contains users, sensors, BSs, and cloud/edge servers.}
\label{fig:1}
   \vspace{-0.5cm}
\end{figure*}

Recent LLM-based multi-agent work enables HAPs coordination and adaptation to natural-language events \cite{han2025agent}, but mainly optimizes platform positioning and user coverage. In contrast, Agentic AI here retrieves domain knowledge, invokes quantitative tools, and decomposes missions into communication, computing, and energy subtasks. Persistent solar power and on-board processing make HAPs suitable carriers for such agents.  In other words, Agentic AI and HAPs are mutually enabling: solar-powered persistence allows the agent to observe and learn over long time horizons, while Agentic AI ensures that harvested energy is converted into useful and sustainable network service.

Motivated by this observation, we develop an Agentic AI-enabled Solar HAP framework for sustainable SAGIN communication and computing. Its main contributions are as follows:
\begin{itemize}
    \item \textbf{A sustainability-oriented HAP-centric SAGIN architecture:} We characterize HAPs as a renewable-energy-powered middle layer between satellites, drones, and terrestrial infrastructure. Rather than treating a HAP only as a relay or aerial base station, the proposed architecture assigns it four integrated roles: regional access point, multi-layer traffic broker, edge-computing hub, and solar-energy-aware service orchestrator.
    \item \textbf{An Agentic AI-native control plane:} We design Agentic AI as an intrinsic component of the HAP operating architecture. The agent jointly observes solar-energy generation, battery condition, flight state, radio links, computing resources, task graphs, and mission priorities. It combines long-term reasoning with link-budget, energy prediction, task scheduling, and optimization tools.
    \item \textbf{Energy-causal communication and computing orchestration:} We formulate a multi-timescale decision process in which the agent controls user association, transmit power, routing, task partitioning, processor frequency, and platform positioning. Decisions are constrained by future energy availability, battery reserves, flight safety, thermal limits, and heterogeneous service-level requirements.
\end{itemize}

\section{Solar-Powered HAP-Centric SAGIN Framework}

\subsection{HAPs as the Sustainable Middle Layer}
The proposed framework contains four interacting platform domains: the space domain, the high-altitude domain, the low-altitude domain, and the terrestrial domain, as shown in Fig. \ref{fig:1}. Satellites provide global reach, timing, remote backhaul, and large-scale observation. Drones provide local mobility, sensing, data collection, and rapidly deployable access. Terrestrial Base Stations (BSs) and edge clouds provide high-capacity processing and stable service where infrastructure remains available. Solar-powered HAPs connect these domains and determine how communication and computation workloads should flow among them.

\begin{table*}[!t]
\centering
\caption{Sustainability-Oriented Comparison of SAGIN Platforms}
\label{tab:sagin_platform_comparison}
\renewcommand{\arrayrulewidth}{0.8pt}
\renewcommand{\tabcolsep}{3.0pt}
{\fontsize{6.8}{8.2}\selectfont
\begin{tabular}{
m{1.2cm}<{\centering}||
m{1.4cm}<{\centering}|
m{1.55cm}<{\centering}|
m{1.75cm}<{\centering}|
m{2.25cm}<{\centering}|
m{2.10cm}<{\centering}|
m{2.35cm}<{\centering}|
m{2.30cm}<{\centering}}
\hline \hline
\rowcolor{gray!15}
\textbf{Platform}
&
\textbf{Mobility}
&
\textbf{Typical Endurance}
&
\textbf{Service Coverage}
&
\textbf{Communication Characteristics}
&
\textbf{Computing Capability}
&
\textbf{Main Sustainability Strength}
&
\textbf{Primary Role in SAGINs}
\\
\hline

\rowcolor{blue!10}
\textbf{Terrestrial BS}
&
Fixed and infrastructure
&
Continuous with grid
&
Local cellular coverage
&
Short links, low latency, and stable topology
&
High and easily upgradeable
&
Efficient where grid power and infrastructure are available
&
High-capacity terrestrial access and cloud-edge support
\\
\hline

\rowcolor{orange!12}
\textbf{Drone}
&
Highly mobile
&
Minutes to hours
&
Local and mission-specific
&
Short links but frequent topology variation
&
Limited by payload and battery capacity
&
Suitable for short-duration and localized missions
&
Local sensing, temporary access, and agile task execution
\\
\hline

\rowcolor{green!10}
\textbf{HAP}
&
Quasi-stationary
&
Weeks to months
&
Persistent wide-area regional coverage
&
Moderate-distance LoS links with low handover frequency
&
Moderate-to-high onboard edge computing
&
Combines renewable operation, long endurance, and wide coverage
&
Regional hub and cross-layer service orchestrator
\\
\hline

\rowcolor{purple!10}
\textbf{Satellite}
&
Fast-moving
&
Several years
&
Very large but moving footprint
&
Long links, higher path loss, and frequent changes
&
Constrained by payload power and maintenance
&
Solar-powered global coverage and long unattended operation
&
Global connectivity and long-distance backhaul
\\
\hline \hline
\end{tabular}}
\vspace{2pt}

\end{table*}

The HAP is equipped with four physical resource subsystems:
\emph{1) A solar-energy subsystem}, including photovoltaic arrays, maximum-power-point tracking, batteries, energy-conversion circuits, and energy/thermal sensors;  \emph{2) A communication subsystem}, including multi-band access and feeder links, beamforming units, software-defined radios, and inter-HAP links; \emph{3) A computing subsystem}, including CPUs, GPUs, memory, storage, and virtualization support for edge applications; \emph{4) A flight subsystem}, including navigation, wind estimation, altitude control, propulsion, and station-keeping functions.

Above these physical resources, the HAP acts as a persistent regional communication and computing hub in SAGINs rather than merely an aerial relay. It can provide wide-area access to ground users; aggregate traffic from drones and sensors; support satellite and terrestrial backhaul; execute delay-sensitive tasks through onboard edge servers; cache frequently requested content and AI models; and coordinate low-altitude platforms when terrestrial infrastructure is unavailable or overloaded \cite{umar2025high}. Its solar-powered and quasi-stationary operation further enables these services to be sustained over long periods with reduced dependence on grid or fuel-based energy. To unify these functions, a HAP-native Agentic AI control plane is embedded into the platform to coordinate energy, communication, computing, and mobility resources according to the current network and mission state. By directly interfacing with the HAP’s sensing, networking, computing, energy-management, and flight-control modules, it transforms the HAP into an autonomous and executable service orchestrator for sustainable SAGIN operation.

Table \ref{tab:sagin_platform_comparison} further emphasizes the uniqueness of HAPs in the middle layer of the SAGIN. However, it is necessary to point out that the platforms are complementary rather than interchangeable. The terrestrial layer should execute workloads requiring very high processing capacity when infrastructure is available. Drones should perform spatially precise sensing and short-term service. Satellites should provide global reach and backbone connectivity. The HAP should handle persistent regional access, cross-layer task aggregation, edge inference, caching, and traffic steering. This division reduces unnecessary long-distance transmissions, avoids excessive drone propulsion and battery replacement, and limits the need for permanent terrestrial densification.

\subsection{Sustainability Through Joint Communication and Computing}

The sustainability gain of a HAP does not arise solely from replacing grid energy with solar energy. It arises from changing how SAGIN tasks are executed.

First, a HAP can terminate communication traffic closer to ground and aerial users than a satellite, reducing path loss and terminal transmission power. Second, it can process raw sensing data before forwarding it through the satellite feeder link. For example, instead of transmitting an entire high-resolution drone video stream, the HAP can execute object detection and forward only detected events, feature vectors, or compressed semantic information \cite{nguyen2025contemporary}. Third, it can cache frequently requested models and content, reducing repeated satellite and terrestrial backhaul transmissions. Fourth, its persistent regional presence enables multiple drones and sensors to share a common computing and communication infrastructure rather than carrying redundant high-performance processors.

However, the HAP cannot indiscriminately accept all tasks. Computation competes with radio transmission and station keeping for harvested energy. The architecture therefore exposes energy state as a first-class network resource. Each communication or computation request is associated with not only latency, reliability, and processing requirements, but also an energy profile and execution deadline \cite{younis2024energy}. The HAP agent admits, delays, partitions, migrates, or rejects tasks according to their mission value and their effect on long-term energy neutrality.

\section{Agentic AI-enabled HAPs}
\subsection{Why the Agent Should Be Native to HAPs}

Prior generative-AI-assisted HAP design couples propulsion modeling with communication beamforming, but employs the agent mainly as an offline research assistant \cite{xing2026generative}. In contrast, our agent is embedded in the operational control loop and jointly observes energy, flight, communication, computing, and mission states. It therefore coordinates online actions across these domains, rather than only assisting model construction or parameter preparation. For example, higher transmit power improves links but reduces the night energy reserve, while additional computing relieves backhaul congestion but increases energy and thermal loads.

A HAP-native agent has access to solar generation, battery status, wind and position, link quality, traffic demand, task queues, processor utilization, and mission priorities. This enables it to distinguish whether service degradation is caused by radio interference, backhaul congestion, computing overload, energy shortage, or platform drift, and to coordinate the corresponding communication, computing, and energy. By directly interfacing with the radio controller, edge runtime, energy-management unit, navigation system, and flight-control interfaces, the agent becomes part of the HAP’s operating intelligence rather than an external advisory tool.

\subsection{The Perception-Reasoning-Decision-Action-Feedback Loop}
The closed-loop operation process is shown in Fig. \ref{fig:2}, and the details are as follows:

 \begin{figure*}[!t]
\centering
 \includegraphics[width=5 in]{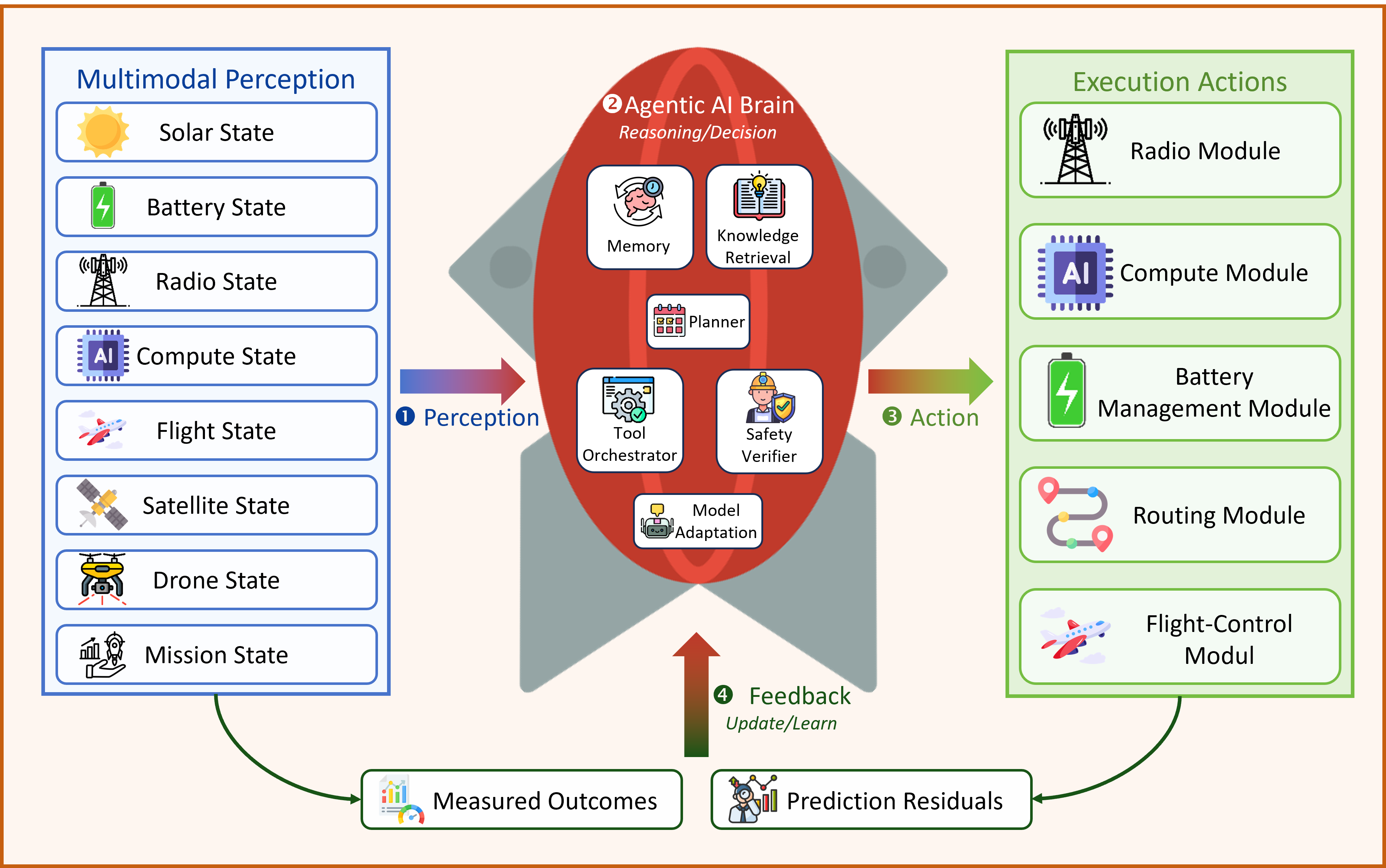}
   \caption{Closed Loop for Agentic AI-Native HAP. It contains the Agentic AI brain, multimodal perception, executable actions, and feedback.}
\label{fig:2}
   \vspace{-0.5cm}
\end{figure*}

\subsubsection{Perception: Constructing a Cross-Domain HAP State} Perception must extend beyond conventional network telemetry. The agent jointly senses five classes of information.
\emph{Energy perception} includes instantaneous solar power, predicted solar irradiance, battery state of charge, state of health, charging efficiency, battery temperature, propulsion power, and the minimum reserve required for safe night operation;
\emph{Communication perception} includes channel quality, interference, beam occupancy, traffic demand, packet queues, backhaul capacity, satellite visibility, predicted satellite handovers, and inter-HAP link conditions;
\emph{Computing perception} includes processor utilization, accelerator availability, memory and storage, task queues, model-loading time, inference accuracy, thermal state, and the energy-per-cycle characteristics of available processors.
Mobility and environment perception include HAP drift, wind speed and direction, cloud and storm conditions below the platform, drone trajectories, user mobility, and the position and velocity of relevant satellites;
\emph{Mission perception} includes task deadlines, reliability classes, semantic importance, data-sovereignty constraints, emergency priorities, and operator policies. This information allows the agent to differentiate, for example, a life-critical rescue-control packet from delay-tolerant environmental data.

\subsubsection{Tool-Augmented Reasoning} Agentic reasoning should not rely on an LLM to perform numerical communication and energy calculations internally. The LLM acts as the planner and tool orchestrator, while quantitative tools supply verifiable results \cite{luo2025toward}.
A typical reasoning sequence is:

\begin{itemize}
    \item \textbf{Identify the active objective:} Determine whether the immediate priority is emergency coverage, task completion, battery preservation, feeder-link congestion relief, or a combination of objectives.
    \item \textbf{Diagnose the dominant bottleneck:} Separate radio-access delay, satellite-backhaul delay, compute-queue delay, energy throttling, and mobility-induced degradation.
    \item \textbf{Forecast future states:} Estimate solar generation, nighttime energy demand, traffic evolution, UAV movement, and satellite-link availability over the relevant planning horizon.
    \item \textbf{Generate candidate plans:} Examples include local HAP execution, drone-HAP split inference, satellite offloading, terrestrial-cloud migration, model compression, task postponement, or HAP repositioning.
    \item \textbf{Quantify each candidate:} Invoke tools to estimate link capacity, execution latency, task accuracy, energy expenditure, battery degradation, and handover overhead.
    \item \textbf{Check feasibility and risk:} Eliminate plans that violate spectrum, computing, thermal, battery-reserve, airspace, or flight-safety constraints.
    \item \textbf{Select and explain a plan:} Choose the plan with the highest long-term mission utility and retain its assumptions and tool outputs for later verification.
\end{itemize}

Meanwhile, the HAP agent should have access to at least six tool classes \cite{liu2026unleashing}: \emph{A solar-wind forecasting tool} for estimating renewable generation and station-keeping energy;
\emph{A satellite ephemeris and handover predictor} for estimating future feeder-link availability;
\emph{A link-budget and beam simulator} for evaluating access, inter-HAP, and satellite links;
\emph{A task-graph and edge-computing scheduler} for estimating execution time, intermediate-data volume, and processor energy;
\emph{A SAGIN digital twin} for testing coordinated radio, computing, energy, and mobility actions before deployment;
\emph{A constrained optimization solver} for refining numerical parameters after the agent has selected a high-level strategy.
Tool use is not an optional enhancement. It is the mechanism that converts semantic agent reasoning into quantitatively defensible network decisions.

\subsection{Multi-Timescale Decision Making} The agent operates over three control horizons.
At the \emph{fast timescale}, ranging from milliseconds to seconds, it controls user scheduling, beam selection, transmit power, spectrum blocks, packet duplication, processor frequency, and task-queue priority;
At the \emph{medium timescale}, ranging from seconds to minutes, it determines task partitioning, model selection, model quantization, caching, inter-layer routing, satellite or terrestrial backhaul selection, and drone-HAP association;
At the \emph{slow timescale}, ranging from minutes to hours, it reserves energy for night operation and migrates noncritical workloads to periods of high solar generation. It also adjusts the HAP service footprint, coordinates station keeping, manages battery degradation, and negotiates workload sharing with neighboring HAPs or satellites.
This hierarchy prevents a short-term radio controller from consuming energy that the flight controller will require later. It also prevents a slow LLM-based planner from being placed in a sub-millisecond scheduling loop for which it is unsuitable.

\subsection{Physical and Network Actions}
The output of the agent is an executable cross-domain plan. \emph{Communication actions} include beam steering, transmit-power control, user association, handover, packet duplication, network coding, and routing; \emph{Computing actions} include task admission, CPU/GPU allocation, dynamic voltage and frequency scaling, model splitting, early-exit inference, precision adjustment, caching, and workload migration; \emph{Energy actions} include charging and discharging policies, payload sleep modes, postponement of noncritical computation, and reservation of energy for station keeping. \emph{Mobility actions} include bounded altitude adjustment, service-area repositioning, and coordination with drone trajectories.
Each action must satisfy an explicit operational requirement. A communication action should improve useful service without creating unacceptable interference. A computing action should meet task latency and accuracy requirements. An energy action should preserve the minimum future battery reserve. A mobility action should remain inside the permitted flight corridor and should provide a net system benefit after propulsion energy is included.

\subsection{Feedback, Reflection, and Evolution}
After execution, the agent compares the predicted and observed outcomes. For example, it determines whether an expected beam gain was achieved, whether task execution consumed more energy than estimated.
Additionally, feedback serves three purposes \cite{zhao2026agentification}. First, it supports immediate correction when the selected action fails. Second, it calibrates link, traffic, computing, and energy models. Third, it creates episodic memory describing the context, reasoning steps, tools used, action, and outcome. When a similar condition occurs in a future day–night cycle, the agent can retrieve the earlier episode instead of solving the entire problem from scratch.


\section{Case Study: Agentic HAP-Assisted Sustainable Disaster Recovery}

\subsection{Experimental Scenario}
We investigate a post-earthquake scenario where most terrestrial BSs and fiber backhauls fail across a 20 $\times$ 20 km$^2$ area. As shown in Fig. \ref{fig:3}, a solar-powered HAP at 20 km altitude provides regional access, edge computing, and traffic aggregation; six drones at 150-300 m collect visual data and cover severely damaged zones. A satellite connects the HAP to the remote cloud, with one surviving terrestrial edge server intermittently accessible at the network boundary. The system serves 120 ground devices with three traffic classes: latency-critical rescue robot control, computation-intensive drone detection tasks, and delay-tolerant sensor monitoring. The HAP supports task processing, splitting, offloading, and deferral, with joint control over radio, computing, and service priorities.
The 60-minute simulation includes three overlapping disturbances: satellite backhaul congestion (15–30 min), concurrent traffic surge (25–45 min), and declining solar generation (after 35 min). The core objective is to sustain critical services and maintain sufficient battery reserve under constrained backhaul and rising computation demand.

\subsection{Experimental Setup}
We adopt an NS-3 and Python co-simulation framework. NS-3 models the SAGIN topology and communication links, while Python implements the proposed Agentic controller and baselines.
Key parameters are configured as follows. HAP-ground/drone links operate at 28 GHz with 400 MHz bandwidth and 40 dBm maximum transmit power. Satellite feeder link capacity drops from 800 to 200 Mbit/s during congestion. The HAP has 550 TOPS\footnote{https://defense-solutions.curtisswright.com/capabilities/open-architectures/other/vpx} on-board computing capacity.  Rescue-control packets are 64 bytes, generated every 10 ms with a 20 ms deadline; visual-inference tasks require 12 Mbit input and 8 Gcycles computation, with a 300 ms deadline; monitoring tasks contain 2 Mbit data and 1 Gcycle computation, with a 3 s deadline.
The extended NS-3 energy module models the HAP energy system: 80 kWh battery (25\% safety reserve), 30 kW peak solar harvesting, 6 kW station-keeping power, 0.2-1.5 kW communication power, and 0.3-2.5 kW computing power. The initial battery State of Charge (SoC) is set to 31\%, representing a stressed condition close to the safety reserve. 
NS-3 reports aggregated states to the Python agent every 30 s for high-level orchestration, while deterministic lower-layer controllers handle millisecond-level scheduling and deadline enforcement. Agent-proposed actions are activated only after battery, computation, radio, and deadline checks.
Experiments run on a PC with an Intel Core i9-14900HX (2.2 GHz) CPU and NVIDIA RTX 4070 (8 GB) GPU. All results are averaged over 50 random seeds with 95\% confidence intervals. Three baselines are included for comparison: an LLM scheme, Proximal Policy Optimization (PPO) \cite{sun2024proportional}, and Multi-Agent PPO (MAPPO) \cite{yao2025multi}. The proposed agent and standalone LLM use the same Qwen2.5-7B model with 4-bit weight quantization and observations, while only the agent invokes forecasting, scheduling, and constraint-verification tools.

\subsection{Evaluation Results}
 \begin{figure*}[!t]
\centering
 \includegraphics[width=7 in]{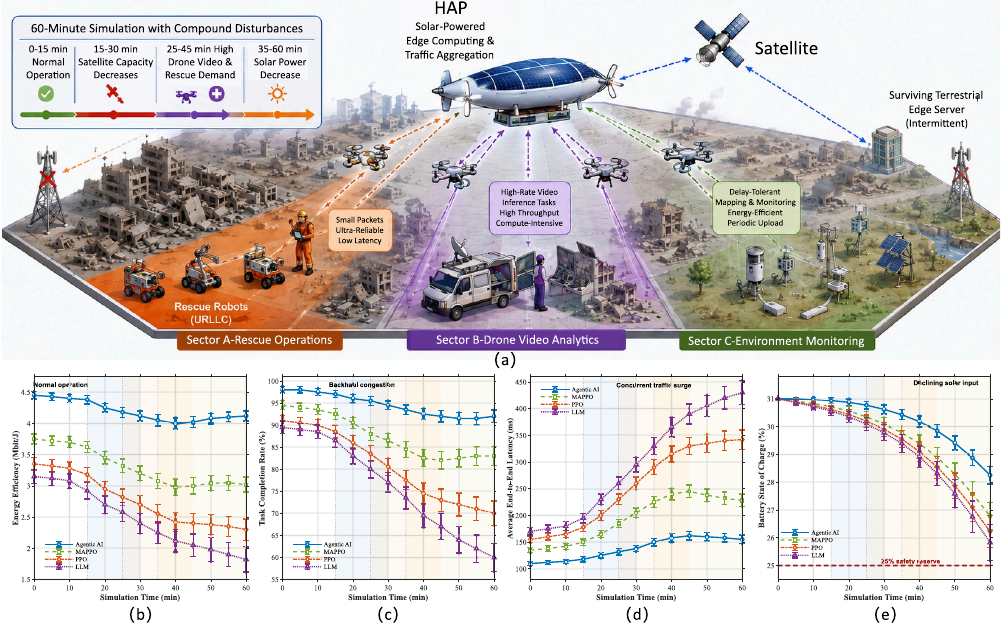}
   \caption{(a) Disaster scenario; Performance of Agentic HAP-assisted sustainable disaster recovery under compound disturbances: (b) energy efficiency; (c) task completion rate; (d) average end-to-end task latency; and (e) battery SoC.}
\label{fig:3}
   \vspace{-0.5cm}
\end{figure*}

We compare the proposed HAP-native Agentic AI with MAPPO, PPO, and an LLM controller under the compound disturbances described above. The shaded intervals indicate satellite-backhaul congestion, the concurrent traffic surge, and declining solar input, respectively. 

As shown in Fig. \ref{fig:3} (b), the proposed scheme maintains an average energy efficiency of 4.20 Mbit/J, outperforming MAPPO, PPO, and the standalone LLM by 27.1\%, 51.6\%, and 70.2\%, respectively. The gap widens when traffic surges and solar decline overlap. The agent interprets rescue intent, selects semantic results instead of raw video, and coordinates local inference, backhaul use, and task deferral. These actions avoid unnecessary feeder-link transmission and high-frequency computation. In contrast, MAPPO and PPO retain policies learned from predefined operating distributions, while the standalone LLM lacks quantitative energy prediction and action verification.
Fig. \ref{fig:3} (c) shows that the proposed scheme achieves an average task completion rate of 94.6\%, compared with 87.5\% for MAPPO, 80.6\% for PPO, and 76.1\% for the standalone LLM. During the compound-stress period, the proposed method retains a completion rate above 92\%, whereas the baselines experience sharper degradation. This improvement results from operator/environment-intent-aware task admission and prioritization. The agent reserves radio and computing resources for rescue-control and victim-detection tasks, while splitting, compressing, or deferring less urgent monitoring workloads. MAPPO coordinates multiple nodes better than PPO but cannot explicitly reinterpret service intent when backhaul and renewable-energy conditions change simultaneously.
Fig. \ref{fig:3} (d) presents the average end-to-end task latency. The proposed method achieves an overall average of 137.8 ms, reducing latency by 29.1\% relative to MAPPO. When the satellite feeder link becomes congested, the standalone LLM and reinforcement-learning baselines continue to offload excessive data or respond only after queues have accumulated. In contrast, the Agentic AI invokes the backhaul estimator and task scheduler before execution, increases local HAP processing, and forwards compact task results. Its latency rises moderately during the traffic peak but stabilizes as the agent replans radio, computing, and offloading decisions. 
As shown in Fig. \ref{fig:3} (e), the battery SoC decreases from 31\% as solar input declines, but the proposed scheme retains 28.25\% at the end of the 60-min simulation. This leaves a 3.25-percentage-point margin above the 25\% safety reserve, whereas the corresponding margins of the three baselines are only 1.78, 1.22, and 0.83 percentage points. The agent preserves this margin by reducing noncritical radio and computing loads and deferring delay-tolerant tasks after the solar decline. It therefore sustains critical services without exhausting the energy reserved for subsequent operation.
These gains arise from closed-loop intent, semantic, and energy-aware coordination rather than isolated resource optimization.

\subsection{Lessons Learned}
Agentic AI should coordinate cross-domain decisions at medium and slow timescales, while conventional controllers handle fast radio and computing actions. Its decisions must consider predicted renewable generation and future battery reserves, and be verified through quantitative tools. Under compound disturbances, mission-aware prioritization, selective local processing, and workload deferral are essential for sustaining critical services and HAP operation.

\section{Future Directions}
\subsection{Trustworthy and Verifiable Agentic Control}
Future HAP agents must provide reliable decisions under uncertain solar generation, dynamic channels, incomplete observations, and unexpected mission changes. Promising solutions include uncertainty-aware reasoning, formal constraint verification, runtime safety guards, and human-overridable control. These mechanisms are essential for preventing unsafe actions such as excessive battery discharge, infeasible task admission, or flight-control conflicts.
\subsection{Collaborative Multi-HAP and Multi-Agent Orchestration}
Large-scale SAGINs will require multiple HAPs, satellites, drones, and terrestrial edge nodes to cooperate rather than operate independently. Future research should investigate distributed agent coordination, task and energy sharing, inter-HAP workload migration, and conflict resolution among heterogeneous agents. A key challenge is achieving global service efficiency while limiting signaling overhead and preserving local autonomy.
\subsection{Digital-Twin-Assisted Lifelong Adaptation}
A HAP agent should continuously adapt to seasonal solar variation, battery aging, changing traffic patterns, new devices, and evolving mission requirements. Integrating high-fidelity SAGIN digital twins with online learning, episodic memory, and sim-to-real adaptation can allow candidate actions to be evaluated before physical execution. This direction can improve long-term robustness while reducing the risks and costs of real-world exploration.
\section{Conclusion}
Solar-powered HAPs can become a sustainable regional middle layer for SAGINs by combining wide-area coverage, renewable-energy harvesting, and onboard computing. This work has presented a HAP-native Agentic AI framework that treats communication, computing, energy, mobility, and mission requirements as a coupled control problem. Rather than acting as a standalone adviser, the agent is embedded in the HAP control plane. Thus, it can perceive cross-domain states; invoke forecasting and optimization tools; verify candidate plans; and execute coordinated actions. Its multi-timescale architecture preserves fast conventional control for radio and scheduling while using Agentic reasoning for long-term battery planning.

The disaster-recovery case study demonstrates the value of this coordination under overlapping backhaul congestion, traffic bursts, and declining solar input. The proposed framework achieves better energy efficiency, task completion, and latency than standalone LLM, PPO, and MAPPO baselines. These gains arise from cross-domain reasoning rather than isolated resource optimization. Moving from simulation to deployment will require trustworthy runtime verification, scalable cooperation among multiple HAPs and heterogeneous agents, and digital-twin-assisted lifelong adaptation. Progress in these areas can transform solar-powered HAPs from persistent aerial infrastructure into safe, autonomous, and energy-aware intelligence hubs for resilient SAGIN services.


\bibliographystyle{IEEEtran}
\bibliography{IEEEabrv,mylib}



\vspace{3em}

\vfill

\end{document}